# Designing Nanostructures for Interfacial Phonon Transport via Bayesian Optimization

Shenghong Ju[1,2], Takuma Shiga[1,2], Lei Feng[1], Zhufeng Hou[2], Koji Tsuda[2,3], Junichiro Shiomi[1,2,*]

[1]Department of Mechanical Engineering, The University of Tokyo, 7-3-1 Hongo, Bunkyo, Tokyo 113-8656, Japan

[2]Center for Materials research by Information Integration, National Institute for Materials Science, 1-2-1 Sengen, Tsukuba, Ibaraki 305-0047, Japan

[3]Department of Computational Biology and Medical Sciences, Graduate School of Frontier Sciences, The University of Tokyo, Kashiwa 277-8561, Japan

[*] Corresponding email: shiomi@photon.t.u-tokyo.ac.jp

## ABSTRACT

We demonstrate optimization of thermal conductance across nanostructures by developing a method combining atomistic Green's function and Bayesian optimization. With an aim to minimize and maximize the interfacial thermal conductance (ITC) across Si-Si and Si-Ge interfaces by means of Si/Ge composite interfacial structure, the method identifies the optimal structures from calculations of only a few percent of the entire candidates (over 60,000 structures). The obtained optimal interfacial structures are non-intuitive and impacting: the minimum-ITC structure is an aperiodic superlattice that realizes 50% reduction from the best periodic superlattice. The physical mechanism of the minimum ITC can be understood in terms of crossover of the two effects on phonon transport: as the layer thickness in superlattice increases, the impact of Fabry–Pérot interference increases, and the rate of reflection at the layer-interfaces decreases. Aperiodic superlattice with spatial variation in the layer thickness has a degree of freedom to realize optimal balance between the above two competing mechanism. Furthermore, aperiodicity breaks the constructive phonon interference between the interfaces inhibiting the coherent phonon transport. The present work shows the effectiveness and advantage of material informatics in designing nanostructures to control heat conduction, which can be extended to other interfacial structures.

**Subject Areas:** Condensed Matter Physics, Computational Physics, Materials Science

# I. INTRODUCTION

Exploration and design of materials with desired thermal transport properties hold importance in thermal management applications such as thermal interface materials [1], heat pipe [2,3], thermoelectrics [4], and thermal insulator [5]. As the length scales of materials are reduced to nanoscale, phonon transport becomes more ballistic (or quasi ballistic) and the interfacial thermal conductance (ITC) determines the heat conduction through the entire material [6,7]. In other words, the heat conduction becomes controllable through manipulating the interface structure. Various individual factors for tuning the ITC have been reported, such as roughness [8,9], vacancy defects [10], lattice orientation [11,12], nanoinclusions [13], and interfacial adhesion or bonding [14,15]. However, these factors are usually coupled with each other and sensitive to the detailed atomic configurations, which makes it hard to identify the total controllability of ITC. The search for optimal structure becomes even more difficult in case of structures with multiple interfaces such as superlattices [16-19], nanocrystals [20], nanocomposites [21,22], where the constructive and deconstructive phonon interference and resonance effects make the heat transport more complicated. The key next-generation technology here can be the materials informatics (MI) [23-25]: integration of material property calculations/measurements and informatics to accelerate the material discovery and design.

During the past decade, informatics has been successfully applied in designing drugs [26,27], polymers [28] and grain boundaries [29], and even in guiding experiments [30]. In the field of heat transfer, MI has been applied to search for thermoelectric materials with low thermal conductivity from developed material database [31,32]. Optimal distribution of nanoparticle size for minimum thermal conductivity was also performed by using a kind of evolutionary algorithm [33]. However, the nanostructure optimization for thermal transport is still in its infancy. Developing effective optimization method for designing nanostructures is necessary and has large potential for application. In this work, we develop a framework by means of atomistic Green's function (AGF) [8,34-36] and Bayesian optimization methods [37], and demonstrate the efficiency and ability to identify non-trivial interfacial structures that realize maximum and minimum ITC.

## II. METHODOLOGY

We explain the basis of the method by taking a problem to design the interfacial alloy structure to tune heat conduction across the Si-Si and Si-Ge interfaces, as shown in Fig. 1 (a) ~ (d). The cross sectional area of the calculated system is 5.43 Å × 5.43 Å and the periodic boundary condition is applied to simulate an interface with infinite cross section. The thickness of the interfacial structure is fixed at 10.86 Å. The interfacial structure between the leads in the simulation cell consists of 16 atoms, which are either Si or Ge. Here, for simplicity, we constrain the fraction of Si and Ge to be 50%. Thus the optimization problem becomes how to arrange the Si and Ge atoms to obtain the largest and smallest ITC.

Four basic elements are required when conducting material informatics, the descriptor, evaluator, calculator, and optimization method. The descriptors are used to describe the possible structure candidates considered during the optimization. In this study, we use a binary flag to describe the state of each atom: '1' and '0' represent Ge and Si atom, respectively. Here, the total number of possible candidates involved in this optimization problem is $C_{16}^8 =12,870$. As for the evaluator, the ITC is chosen to quantitatively evaluate the performance of each configuration. The AGF method [8,34-36] was employed to calculate the phonon transmission function, $\Xi(\omega)$,

$$\Xi(\omega) = \text{Trace}[\Gamma_L G^r \Gamma_R G^a], \qquad (1)$$

where $\omega$ is the phonon frequency, $G^r$ and $G^a$ are the retarded and advanced Green's functions of scattering region, the level broadening matrices $\Gamma_L = i(\Sigma_L^r - \Sigma_L^a)$ and $\Gamma_R = i(\Sigma_L^r - \Sigma_L^a)$ describe the rates of inflow from left lead and outflow into right lead, and $\Sigma_L$ and $\Sigma_R$ are the self-energies, which are calculated from surface Green's functions of left and right leads, respectively. With the two leads kept at different temperatures $T_L$ and $T_R$, the heat current flows through the device is given by the Landauer formula [38],

$$J = \frac{1}{2\pi} \int \hbar\omega[f_L(\omega,T_L) - f_R(\omega,T_R)]\Xi(\omega)d\omega, \qquad (2)$$

where $f_L$ and $f_R$ are the Bose-Einstein distributions of phonons. In the limit of small temperature differences, the value of ITC ($\sigma$) at average temperature $T$ can be further obtained by,

$$\sigma = \frac{\hbar^2}{2\pi k_B T^2 S} \int \omega^2 \Xi(\omega) \frac{e^{\hbar\omega/k_B T}}{\left(e^{\hbar\omega/k_B T} - 1\right)^2} d\omega, \qquad (3)$$

where $k_B$ is the Boltzmann constant, $S$ is the cross sectional area. In this work, the AGF calculation was conducted by using Atomistix ToolKit simulation package (ATK) [39] with Tersoff potentials [40,41]. The number of transverse $k$ points within the Brillouin zone perpendicular to the phonon transport direction is selected as 20×20, which has been tested to ensure convergence of the transmission calculation.

We employed our developed open-source Bayesian optimization library COMBO [37] to perform the optimization process. Bayesian optimization is an experimental design algorithm based on machine learning [42]. Suppose that ITC of $n$ candidates are already calculated, and we are to choose the next one to calculate. A Bayesian regression function is learned from n pairs of descriptors and ITCs (i.e., training examples). For each of the remaining candidates, a predictive distribution of ITC is estimated. The best candidate is chosen based on the criterion of expected improvement [42]. Finally, ITC is calculated for the chosen candidate, and it is added to the training examples. By repeating this procedure, the calculation of ITC is scheduled optimally, and the best candidate can be found quickly.

As the prediction model, we employed a Bayesian linear regression model combined with a random feature map,

$$y = \boldsymbol{w}^T \boldsymbol{\phi}(\boldsymbol{x}) + \varepsilon, \tag{4}$$

where $\boldsymbol{x}$ is a $d$-dimensional vector corresponds to a candidate, $\boldsymbol{w}$ is a $D$-dimensional weight vector, $\varepsilon$ is the noise subject to normal distribution with mean 0 and variance $\sigma$. The random feature map is chosen so that the inner product corresponds to the Gaussian kernel [43],

$$\boldsymbol{\phi}(\boldsymbol{x})^T \boldsymbol{\phi}(\boldsymbol{x}^{\cdot}) = \exp\left(-\frac{\left\|\boldsymbol{x} - \boldsymbol{x}^{\cdot}\right\|^2}{\eta^2}\right). \tag{5}$$

The performance of the model depends on hyperparameters $\sigma$ and $\eta$. In COMBO, they are initialized according a heuristic procedure by Yang et al. [44]. Whenever 20 training examples are added, the hyperparameters are periodically configured and updated by maximizing the type-II likelihood.

## III. RESULS AND DISCUSSIONS

To test the performance of Bayesian optimization, 10 rounds of optimization were conducted with different initial choices of 20 candidates. As shown in Fig. 1 (e) and (f), all optimizations come to convergence within calculations of 438 structures, which is only 3.4% of the total number of candidates (12,870). To check the accuracy of the optimization, the ITC of all candidates were also calculated, and the maximum/minimum ITC and the corresponding structures were confirmed to be exactly the same as those obtained by Bayesian optimization. The probability distribution of ITC shown in the insets of Fig. 1 (e) and (f) show that the probabilities decrease as ITC values approach the minimum and maximum, confirming that the current problem is appropriate for optimization. The ratios of maximum to minimum ITC for Si-Si and Si-Ge interfaces are 2.4 and 2.2, respectively, which indicates that the ITC significantly dependents on the interfacial nanostructure.

Figures 1 (g) and (h) compare the phonon transmission function of optimal structures with those of bare Si-Si and Si-Ge interfaces. Note that the bare Si-Si interface here means the absence of interface. For phonon frequency lower than 3 THz, the transmission of different structures are almost the same because the transport of long wavelength phonons is not sensitive to the interfacial structures with smaller length scales. On the other hand, for higher phonon frequency, the transmission function strongly depends on the structure. The optimal structure with maximum ITC for Si-Si interface shown in Fig. 1 (a) is intuitive as the structure provides continuum path of Si for phonons to coherently propagate. However, the other three optimal structures shown in Fig. 1 (b) ~ (d) are not intuitive. The optimal structure with maximum ITC for Si-Ge interfaces shown in Fig. 1 (b) can be considered as a kind of rough interface, and the phonon transmission are clearly enhanced at frequency range 4~5 THz and 8~11 THz. This agrees with the previous AGF calculation result on rough interface [8], which showed that the roughness can enhance the phonon transmission at interface.

The structures with minimum ITC for both Si-Si and Si-Ge interface shown in Fig.1 (c) and (d) are aperiodic superlattice with the constituent layers perpendicular to the direction of heat conduction. The structure is different from periodic superlattices, which have been widely studied as a class of materials with low thermal conductivity motivated particularly by thermoelectric applications. Their thermal conductivity has been found to take a minimum with respect to the layer thickness due to the crossover between wave interference and particle scattering of phonons at the periodic inter-layer interfaces [17]. As shown in a recent study, aperiodic superlattices can have lower thermal conductivity than the periodic ones [45]. However, as thickness of each layer in superlattice is different, the underlying physics becomes complicated and identifying the optimal structure from vast number of candidates based on physical principles has been a challenge.

Based on the knowledge learnt above, we extend the ITC-minimization problem of interfacial structure to a realistic setup by enlarging the length scales of the structure to the extent controllable in the experiments [46]: the thickness of the unit layer (UL) is 5.43 Å, and total thickness of interfacial structure ranges from 8 to 16 ULs (from 4.35 nm to 8.69 nm). To compensate the increase in the number of candidates, we limit the candidates to Si/Ge superlattices (periodic and aperiodic) based on the above finding that the alloy structure with minimum ITC is a superlattice. Figure 2 illustrates the 8-UL superlattice structure in case of Si-Si interface. Similarly to the descriptors used in the alloy structure optimization, 8 binary flags were used to indicate the state of each UL ('1' indicates Ge and '0' indicates Si). The descriptor for the case shown in Fig. 2 is (10110010).

Table 1 lists all the optimal superlattice structures for Si-Si and Si-Ge interfaces optimized for various total thickness. For each layer thickness, the optimization was performed for equal and variable fraction of Si/Ge atoms. The number of candidates in each case is also listed in brackets. As expected, the numbers of candidates in case of variable fraction are much larger than those with fixed (and equal) fraction. It can be seen that the superlattice always begins or ends with a different material layer from the lead region, which is understandable because it enhances the phonon scattering. Figure 3 (a) summarizes the minimum ITC values versus the number of ULs (i.e. total thickness of the interfacial structure). As the number of UL increases, the minimum ITC decreases. For Si-Si interfaces, the minimum ITC of the variable-fraction case is smaller than that of the fixed fraction case, while for Si-Ge interfaces, the difference between the two cases is very small. Figure 3 (b) compares the ITC of the obtained aperiodic structure at Si-Si interface and with that of traditional periodic [10] superlattices with the best period thickness (optimized separately for each total length). We found that ITC of the designed aperiodic superlattices with fixed Si/Ge fraction is significantly smaller (by 20~50%). Not to mention, making the fraction variable can further reduce the ITC.

Now that the optimal structures have been identified, we look into the mechanisms behind the small ITC. First obvious attempt is to see it from the view of phonon dispersion relations and phonon density of states (DOS), as broadly done to discuss phonon interference. Taking the 10-UL superlattice structures as an example, we choose three typical structures for comparison, the obtained optimal structure (1101010001) and the periodic superlattices with minimum layer thickness (1010101010) and maximum layer thickness (1111100000). Figure 3 (c) and (d) shows the phonon dispersion, phonon DOS and phonon transmission function. Comparing the phonon dispersion relations of structures (1101010001) and (111110000) (Fig. 3 (c)), the difference is quite small even though their ITC values differ by a factor of 2. The DOS of the three structures are also almost the same except for some minor differences in the local peaks. However, we can see obvious difference in the phonon transmission function shown in Fig. 3 (d), and the optimal structure clearly exhibits the minimum transmission. These suggest the incapability of phonon dispersion or DOS in explaining the mechanism of minimum ITC in the optimal structure.

We then take another route by breaking the characteristics of the structure into the thickness of a single layer (layer thickness) and the number of interfaces in the superlattice. For this, we constructed model systems that allow us to independently vary the layer thickness and the number of interfaces as shown in Fig. 4 (a) and (c). In the former, the layer thickness is varied by fixing the number of interfaces, and in the latter, the number of interfaces is varied by fixing the length per layer. Figures 4 (a) and (c) show that as the layer thickness and number of thickness increase the ITC decreases and eventually asymptotically converges to a constant value. The corresponding phonon transmission functions in Fig. 4 (b) and (d) show that the general dependence on the layer thickness and number of thickness follows that of ITC.

The dependence on number of interface is intuitive since the scattering should increase with the number of interface. The reason for the convergence will be discussed later. To understand the dependence on the layer thickness, we have adopted the 1-D atomic chain model and calculated phonon transmission by changing the thickness of scattering region and the distance between two interfaces $d$ (see supplementary materials Fig. S1 and Fig. S2). The results clearly exhibit the Fabry–Pérot oscillations [47,48], with both constructive and destructive phonon interferences, which gives rise to the strong dependence of phonon transmission function on the distance between two interfaces. While, when $d$ is small, the resonance occurs at specific frequencies, as $d$ increases, the number of resonant frequencies increases and eventually covers the entire frequency range. As a result thermal conductance decreases faster when $d$ is small and eventually saturates to a constant value as $d$ increase. This trend resembles that of Si/Ge system in Fig. 4(a), and thus, we attribute the observed trend to the Fabry–Pérot oscillations.

On considering a superlattice with a given total thickness, the layer thickness and number of interfaces are two competitive parameters, and this gives rise to the optimal structure with minimum ITC. On optimizing the balance between layer thickness and number of interfaces, aperiodic superlattice can be superior to the periodic counterpart because it has degree of freedom to spatially distribute parts with different layer thicknesses and numbers of interfaces.

To further highlight the above discussed competition of the two effects, the ITC of all the candidates were calculated for the 14-UL superlattice with fixed (equal) Si/Ge fraction and 10-UL superlattice with variable fraction. The ITC versus the number of interfaces in the superlattices is shown in Fig. 4 (e) and (f). In both cases, the profile of ITC with respect to the number of interfaces takes a minimum, which confirms the competition. Figure 4 (e) and (f) also show that structures with the same number of interfaces can result in significantly different ITC due to the difference in the thickness of the constituent layers.

Now, to gain deeper understanding into the physics of phonon transport in the optimized structure, we look into the role of phonon coherence, which can cause constructive and destructive interferences. To this end, we attempt to separate the transmission due to particle (incoherent) effect and wave (coherent) effect in superlattices by comparing the phonon transmission from full AGF calculation and the cascade transmission model in the view of particle transport [49-51]. In the cascade model, the phonon transmission of each component is assumed to be independent from each other and the effective phonon transmission $\Xi_{cascade}$ is obtained as,

$$\frac{1}{\Xi_{cascade}} = \sum_i \frac{1}{\Xi_i} - \frac{N-1}{\Xi_{Si}}, \qquad (6)$$

where $N$ is the number of components, $\Xi_i$ is the transmission coefficients of $i$th component, and $\Xi_{Si}$ is the phonon transmission of perfect silicon crystal. Figure 5 (a) compares the results obtained by the cascade model and full AGF calculation for periodic superlattice with three different numbers of periods (2, 4, and 6), where a period consists of one Si UL and one Ge UL ('10' in terms of the descriptor). With increasing superlattice periods, the transmission of the full AGF calculation converges quickly, while the transmission of the cascade model keeps decreasing. This is understandable because the cascade model reflects only the incoherent phonon transport while the full AGF calculation captures the coherent phonon transport crossing multiple layers. The convergence of the full AGF transmission to smaller values indicates the contribution from constructive interference in the periodic superlattices. This suggests that there should be a room to further reduce the phonon transmission and ITC if the constructive phonon interference can be broken. Note that these structures are the same as the ones calculated in Fig. 4(c) and thus explains the reason for the convergence.

Take the optimal 10-UL superlattice for phonon transport from Si to Si with fixed and equal Si/Ge fraction as an example, 1101010001 can be divided to four individual components including one '0110' and three '010'. Figure 5 (b) shows the comparison between the effective phonon transmission from the cascade model and the transmission obtained from full AGF calculation. We find that the cascade model reproduces the general trend and magnitude of the transmission functions although there are certainly some differences in the details. The agreement suggests that the constructive phonon interference is suppressed by the aperiodic structure, and the phonon transmission approaches the incoherent phonon-transport limit, leading to the minimum ITC.

**IV. CONCLUSIONS**

In conclusion, we have identified the Si/Ge-composite interfacial structures that minimize/maximize the ITC across Si-Si and Si-Ge interfaces by the developed framework combining atomistic Green's function and Bayesian optimization methods. The optimal structures were obtained by calculating only a few percent of the total candidate structures, considerably saving the computational resources. The validity and capability of the method are demonstrated by identifying the thin interfacial structures with the optimal Si/Ge configurations among all the possible candidates. Based on the finding that the interfacial structures with minimum ITC take a form of aperiodic superlattice, we extended the search to thicker structures (up to 8.69 nm), and identified non-intuitive structures whose ITCs are significantly smaller than those of the optimal periodic superlattices. The small ITC in the aperiodic structures originates from their degree of freedom to mutual-adoptively balance the two competing effects: Fabry–Pérot wave interference and interfacial particle scattering, which reduces ITC as thickness of the constituent layers in superlattice increases and decreases, respectively. As a results, the optimal aperiodic structure was found to restrain the constructive phonon interference, making the phonon transport to approach its incoherent limit. The present work shows the effectiveness and advantage of material informatics in designing nanostructures to control heat conduction.

**ACKNOWLEDGMENTS**

This work was supported by "Materials research by Information Integration" Initiative (MI$^2$I) project of the Support Program for Starting Up Innovation Hub from Japan Science and Technology Agency (JST). This research used computational resources of the HPCI system provided by Information Initiative Center at Hokkaido University and Academic Center for Computing and Media Studies at Kyoto University through the HPCI System Research Project (Project ID: hp160161). This work was performed using facilities of the Institute for Solid State Physics, the University of Tokyo. This work was partially supported by KAKENHI (Grand Nos. 16H04274, 15K17982).

## References

[1] R. Prasher, *Thermal interface materials: Historical perspective, status, and future directions*, Proceedings of the Ieee **94**, 1571 (2006).

[2] A. Faghri, *Review and Advances in Heat Pipe Science and Technology*, Journal of Heat Transfer **134**, 123001 (2012).

[3] H. N. Chaudhry, B. R. Hughes, and S. A. Ghani, *A review of heat pipe systems for heat recovery and renewable energy applications*, Renewable and Sustainable Energy Reviews **16**, 2249 (2012).

[4] Z. T. Tian, S. Lee, and G. Chen, *Heat Transfer in Thermoelectric Materials and Devices*, Journal of Heat Transfer **135**, 061605 (2013).

[5] O. Kaynakli, *A review of the economical and optimum thermal insulation thickness for building applications*, Renewable and Sustainable Energy Reviews **16**, 415 (2012).

[6] S. Volz, J. Shiomi, M. Nomura, and K. Miyazaki, *Heat conduction in nanostructured materials*, Journal of Thermal Science and Technology **11**, JTST0001 (2016).

[7] P. M. Norris, N. Q. Le, and C. H. Baker, *Tuning Phonon Transport: From Interfaces to Nanostructures*, Journal of Heat Transfer **135**, 061604 (2013).

[8] Z. Tian, K. Esfarjani, and G. Chen, *Enhancing phonon transmission across a Si/Ge interface by atomic roughness: First-principles study with the Green's function method*, Physical Review B **86** (2012).

[9] S. Merabia and K. Termentzidis, *Thermal boundary conductance across rough interfaces probed by molecular dynamics*, Physical Review B **89** (2014).

[10] Y. Liu, C. Hu, J. Huang, B. G. Sumpter, and R. Qiao, *Tuning interfacial thermal conductance of graphene embedded in soft materials by vacancy defects*, The Journal of Chemical Physics **142**, 244703 (2015).

[11] P. K. Schelling, S. R. Phillpot, and P. Keblinski, *Kapitza conductance and phonon scattering at grain boundaries by simulation*, Journal of Applied Physics **95**, 6082

(2004).

[12] S.-H. Ju and X.-G. Liang, *Investigation on interfacial thermal resistance and phonon scattering at twist boundary of silicon*, Journal of Applied Physics **113**, 053513 (2013).

[13] M. Sakata, T. Hori, T. Oyake, J. Maire, M. Nomura, and J. Shiomi, *Tuning thermal conductance across sintered silicon interface by local nanostructures*, Nano Energy **13**, 601 (2015).

[14] M. Sakata, T. Oyake, J. Maire, M. Nomura, E. Higurashi, and J. Shiomi, *Thermal conductance of silicon interfaces directly bonded by room-temperature surface activation*, Applied Physics Letters **106**, 081603 (2015).

[15] K. Zheng, F. Sun, X. Tian, J. Zhu, Y. Ma, D. Tang, and F. Wang, *Tuning the Interfacial Thermal Conductance between Polystyrene and Sapphire by Controlling the Interfacial Adhesion*, ACS Applied Materials & Interfaces **7**, 23644 (2015).

[16] M. N. Luckyanova, J. Garg, K. Esfarjani, A. Jandl, M. T. Bulsara, A. J. Schmidt, A. J. Minnich, S. Chen, M. S. Dresselhaus, Z. Ren, E. A. Fitzgerald, and G. Chen, *Coherent Phonon Heat Conduction in Superlattices*, Science **338**, 936 (2012).

[17] J. Garg and G. Chen, *Minimum thermal conductivity in superlattices: A first-principles formalism*, Physical Review B **87** (2013).

[18] Y. Chen, D. Li, J. R. Lukes, Z. Ni, and M. Chen, *Minimum superlattice thermal conductivity from molecular dynamics*, Physical Review B **72** (2005).

[19] J. Ravichandran, A. K. Yadav, R. Cheaito, P. B. Rossen, A. Soukiassian, S. J. Suresha, J. C. Duda, B. M. Foley, C. H. Lee, Y. Zhu, A. W. Lichtenberger, J. E. Moore, D. A. Muller, D. G. Schlom, P. E. Hopkins, A. Majumdar, R. Ramesh, and M. A. Zurbuchen, *Crossover from incoherent to coherent phonon scattering in epitaxial oxide superlattices*, Nature Materials **13**, 168 (2014).

[20] P. Němec and P. Malý, *Temperature dependence of coherent phonon dephasing in $CsPbCl_3$ nanocrystals*, Physical Review B **72** (2005).

[21] X. Li and R. Yang, *Equilibrium molecular dynamics simulations for the thermal conductivity of Si/Ge nanocomposites*, Journal of Applied Physics **113**, 104306 (2013).

[22] S. Ju and X. Liang, *Detecting the phonon interference effect in Si/Ge nanocomposite by wave packets*, Applied Physics Letters **106**, 203107 (2015).

[23] K. Rajan, *Materials Informatics: The Materials "Gene" and Big Data*, Annual Review of Materials Research **45**, 153 (2015).

[24] K. Rajan, *Materials informatics*, Materials Today **15**, 470 (2012).

[25] A. Agrawal and A. Choudhary, *Perspective: Materials informatics and big data: Realization of the "fourth paradigm" of science in materials science*, APL Materials **4**, 053208 (2016).

[26] D. S. Wishart, *Bioinformatics in drug development and assessment*, Drug Metabolism Reviews **37**, 279 (2005).

[27] T. L. Blundell, B. L. Sibanda, R. W. Montalvao, S. Brewerton, V. Chelliah, C. L. Worth, N. J. Harmer, O. Davies, and D. Burke, *Structural biology and bioinformatics in drug design: opportunities and challenges for target identification and lead discovery*, Philosophical Transactions of the Royal Society B: Biological Sciences **361**, 413 (2006).

[28] N. Adams and P. Murray-Rust, *Engineering Polymer Informatics: Towards the Computer-Aided Design of Polymers*, Macromolecular Rapid Communications **29**, 615 (2008).

[29] S. Kiyohara, H. Oda, K. Tsuda, and T. Mizoguchi, *Acceleration of stable interface structure searching using a kriging approach*, Japanese Journal of Applied Physics **55**, 045502 (2016).

[30] P. B. Wigley, P. J. Everitt, A. van den Hengel, J. W. Bastian, M. A. Sooriyabandara, G. D. McDonald, K. S. Hardman, C. D. Quinlivan, P. Manju, C. C. Kuhn, I. R. Petersen, A. N. Luiten, J. J. Hope, N. P. Robins, and M. R. Hush, *Fast machine-learning online optimization of ultra-cold-atom experiments*, Scientific Reports **6**, 25890 (2016).

[31] A. Seko, A. Togo, H. Hayashi, K. Tsuda, L. Chaput, and I. Tanaka, *Prediction of Low-Thermal-Conductivity Compounds with First-Principles Anharmonic Lattice-Dynamics Calculations and Bayesian Optimization*, Physical Review Letters **115**, 205901 (2015).

[32] J. Carrete, W. Li, N. Mingo, S. Wang, and S. Curtarolo, *Finding Unprecedentedly Low-Thermal-Conductivity Half-Heusler Semiconductors via High-Throughput Materials Modeling*, Physical Review X **4**, 011019 (2014).

[33] H. Zhang and A. J. Minnich, *The best nanoparticle size distribution for minimum thermal conductivity*, Scientific Reports **5**, 8995 (2015).

[34] X. Li and R. Yang, *Size-dependent phonon transmission across dissimilar material interfaces*, Journal of Physics: Condensed Matter **24**, 155302 (2012).

[35] W. Zhang, T. S. Fisher, and N. Mingo, *Simulation of Interfacial Phonon Transport in Si–Ge Heterostructures Using an Atomistic Green's Function Method*, Journal of Heat Transfer **129**, 483 (2007).

[36] J. S. Wang, J. Wang, and J. T. Lü, *Quantum thermal transport in nanostructures*, The European Physical Journal B **62**, 381 (2008).

[37] T. Ueno, T. D. Rhone, Z. Hou, T. Mizoguchi, and K. Tsuda, *COMBO: An efficient Bayesian optimization library for materials science*, Materials Discovery, In Press (2016).

[38] R. Landauer, *Electrical Resistance of Disordered One-Dimensional Lattices*, Philosophical Magazine **21**, 863 (1970).

[39] I. M. Khalatnikov, *Theory of the Kapitza temperature discontinuity at a solid body-liquid helium boundary*, Zh. Eksp. Teor. Fiz. **22**, 687 (1952).

[40] J. Tersoff, *Modeling Solid-State Chemistry - Interatomic Potentials for Multicomponent Systems*, Physical Review B **39**, 5566 (1989).

[41] J. Tersoff, *Erratum: Modeling solid-state chemistry: Interatomic potentials for multicomponent systems*, Physical Review B **41**, 3248 (1990).

[42] D. R. Jones, *A taxonomy of global optimization methods based on response surfaces*, Journal of Global Optimization **21**, 345 (2001).

[43] A. Rahimi and B. Recht, in *Advances in Neural Information Processing Systems 20* (NIPS, 2007), pp. 1177.

[44] Z. Yang, A. J. Smola, L. Song, and A. G. Wilson, in *Proceedings of the Eighteenth International Conference on Artificial Intelligence and Statistics* (JMLR: Workshop and Conference Proceedings, San Diego, CA, USA, 2015), pp. 1098.

[45] B. Qiu, G. Chen, and Z. Tian, *Effects of Aperiodicity and Roughness on Coherent Heat Conduction in Superlattices*, Nanoscale and Microscale Thermophysical Engineering **19**, 272 (2015).

[46] H. N. Lee, H. M. Christen, M. F. Chisholm, C. M. Rouleau, and D. H. Lowndes, *Strong polarization enhancement in asymmetric three-component ferroelectric superlattices*, Nature **433**, 395 (2005).

[47] P. Hyldgaard, *Resonant thermal transport in semiconductor barrier structures*, Physical Review B **69** (2004).

[48] P. E. Hopkins, P. M. Norris, M. S. Tsegaye, and A. W. Ghosh, *Extracting phonon thermal conductance across atomic junctions: Nonequilibrium Green's function approach compared to semiclassical methods*, Journal of Applied Physics **106**, 063503 (2009).

[49] J. Wang, L. Li, and J.-S. Wang, *Tuning thermal transport in nanotubes with topological defects*, Applied Physics Letters **99**, 091905 (2011).

[50] S. Datta, *Quantum Transport: Atom to Transistor* (Cambridge University Press, 2005).

[51] S. Datta, *Electronic Transport in Mesoscopic Systems* (Cambridge University Press, 1995).

**FIGURES, TABLES AND CAPTIONS**

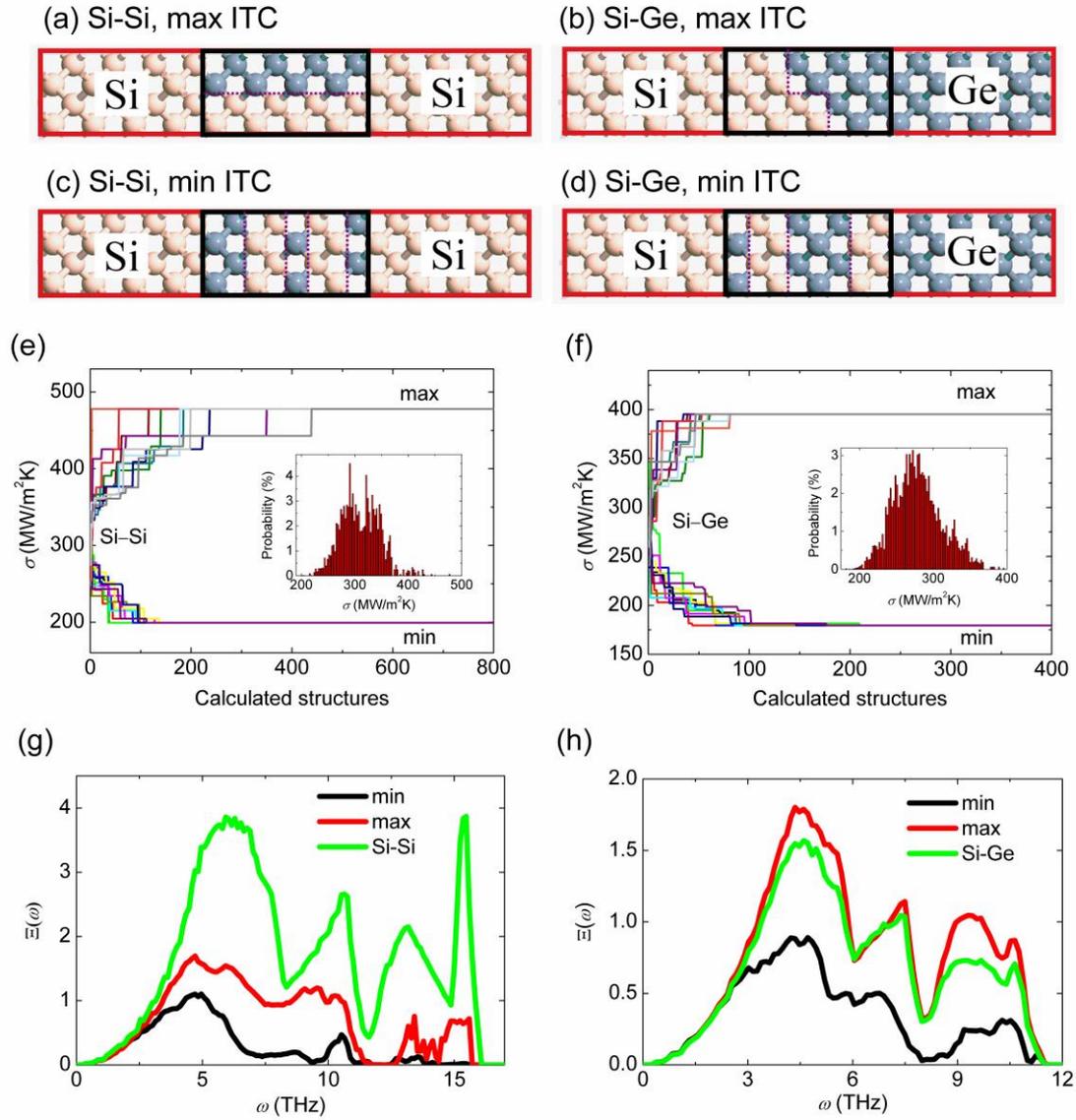

**Fig. 1** Interfacial Si/Ge alloy structure optimization. (a)~(d) Optimal structures with the maximum and minimum interfacial thermal conductance (ITC) for Si-Si and Si-Ge interface. (e) and (f) shows the 10 optimization runs with different initial choices of candidates, where the insets show the probability distribution of ITC. (g) and (f) compare the phonon transmission functions of the optimized interfacial structure (maximum and minimum ITC) and bare Si-Si and Si-Ge interfaces (i.e. no interfacial structure).

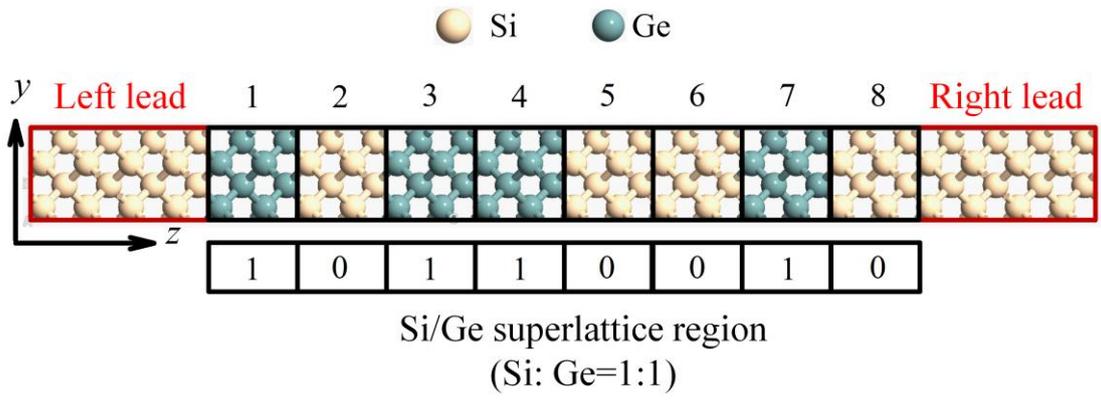

**Fig. 2** Sketch of the 8-UL superlattice structure.

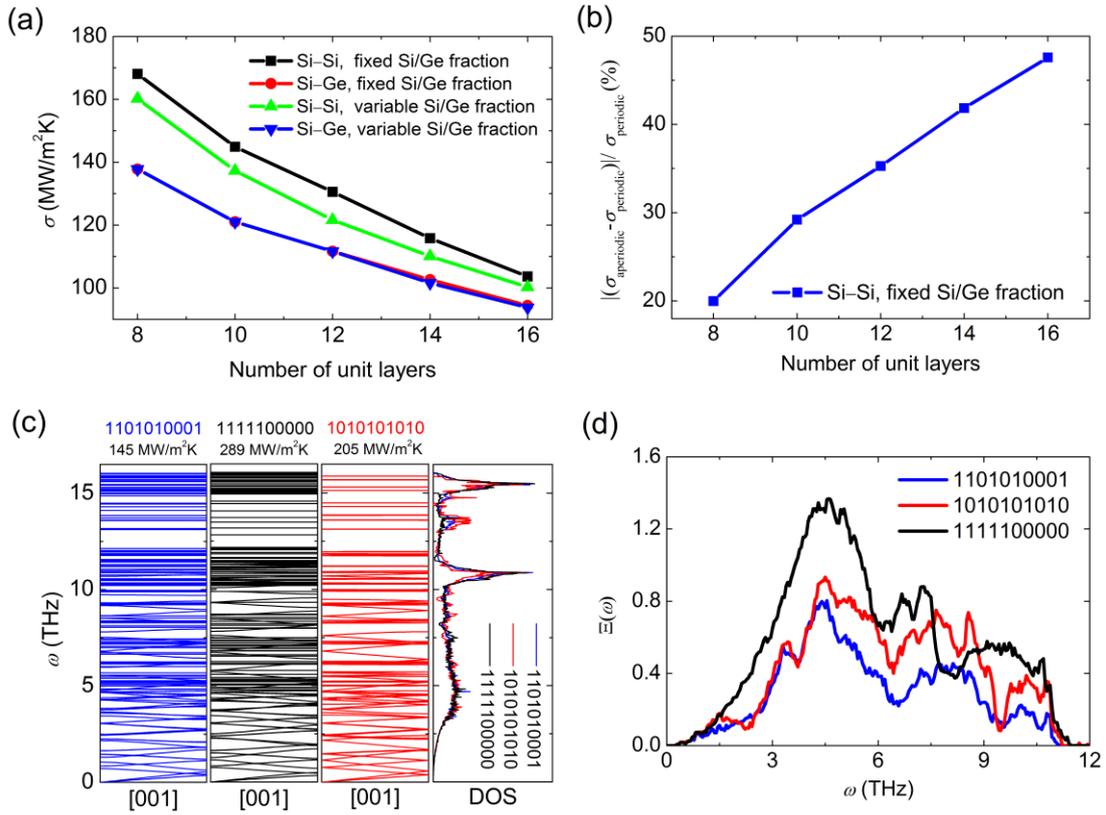

**Fig. 3** (a) Comparison of ITC of interfacial structures at Si-Si and Si-Ge interaces with variable ULs. (b) The reductoin ratio |($\sigma_{\text{aperiodic}}$-$\sigma_{\text{periodic}}$)|/ $\sigma_{\text{periodic}}$, where $\sigma_{\text{aperiodic}}$ is the ITC of the optimized aperiodic superlattice, and $\sigma_{\text{periodic}}$ is the ITC of the corresponding periodic superlattice with period thickness optimized for each number of ULs (i.e. total length). The comparision of (c) the phonon dispersion and DOS, and (d) the phonon transmission for three superlattice structures optimal aperiodic superlattice (1101010001), and periodic superlattice with the largest (1111100000) and smallest (1010101010) periodic thickness.

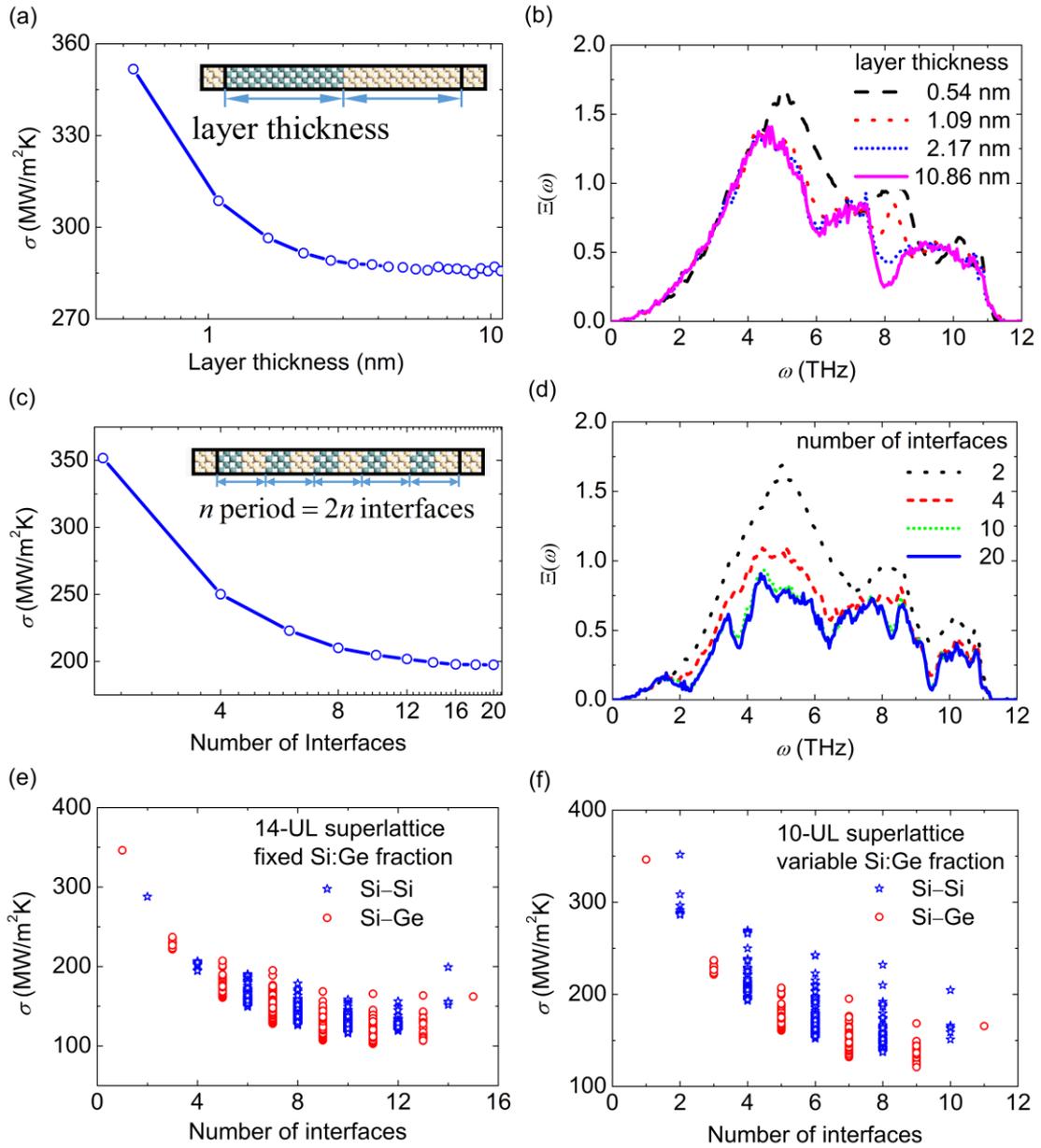

**Fig. 4** (a) and (b) ITC and phonon transmission versus the layer thickness. (c) and (d) ITC and phonon transmission versus the number of interfaces. (e) and (f) ITC versus number of interfaces for cases of 14-UL superlattice with equal Si/Ge layer number and 10-UL superlattice with variable Si/Ge fraction.

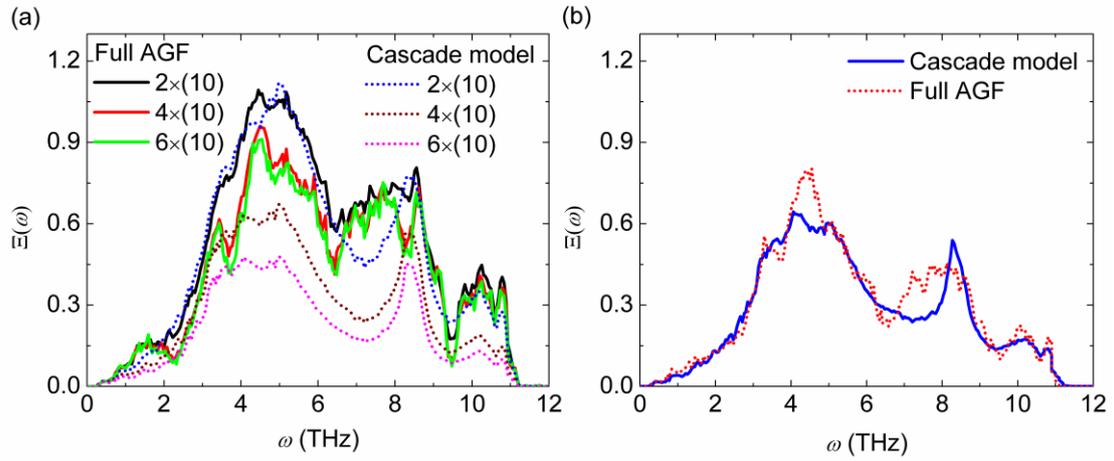

**Fig. 5** Comparision between the phonon transmission obtained from the cascade model and full AGF calculation. (a) periodic '10' superlattices with different number of periods (b) 10-UL optimal structure.

**Table 1** Optimal superlattice structure obtained through Bayesian optimization for different setups and constraints. The numbers in the brackets are the number of candidates in each case.

| UL number | Si-Si fixed Si/Ge fraction | Si-Si variable Si/Ge fraction | Si-Ge fixed Si/Ge fraction | Si-Ge variable Si/Ge fraction |
|---|---|---|---|---|
| 8 | 11000101 (70) | 11101101 (256) | 10100110 (70) | 10100110 (256) |
| 10 | 1101010001 (252) | 1110110101 (1024) | 100010110 (252) | 100010110 (1024) |
| 12 | 101100100101 (924) | 110110101001 (4096) | 100101010110 (924) | 100101010110 (4096) |
| 14 | 11011000101001 (3432) | 11001010110111 (16384) | 10011001010110 (3432) | 10010101101110 (16384) |
| 16 | 1100010010110101 (12870) | 1100101110110101 (65536) | 1010110110010010 (12870) | 1001010101101110 (65536) |

Shenghong Ju[1,2], Takuma Shiga[1,2], Lei Feng[1], Zhufeng Hou[2], Koji Tsuda[2,3], Junichiro Shiomi[1,2,*]

[1]Department of Mechanical Engineering, The University of Tokyo, 7-3-1 Hongo, Bunkyo, Tokyo 113-8656, Japan

[2]Center for Materials research by Information Integration, National Institute for Materials Science, 1-2-1 Sengen, Tsukuba, Ibaraki 305-0047, Japan

[3]Department of Computational Biology and Medical Sciences, Graduate School of Frontier Sciences, The University of Tokyo, Kashiwa 277-8561, Japan

[*]Corresponding email: shiomi@photon.t.u-tokyo.ac.jp

Here we present the Fabry–Pérot oscillations of phonon transmission in one dimensional chain by changing the scattering layer thickness (Fig. S1) and the distance between two interfaces (Fig. S2). For both of the two cases, the normalized integration of phonon transmission convergent to almost constant value, even though the phonon transmissions are quite different.

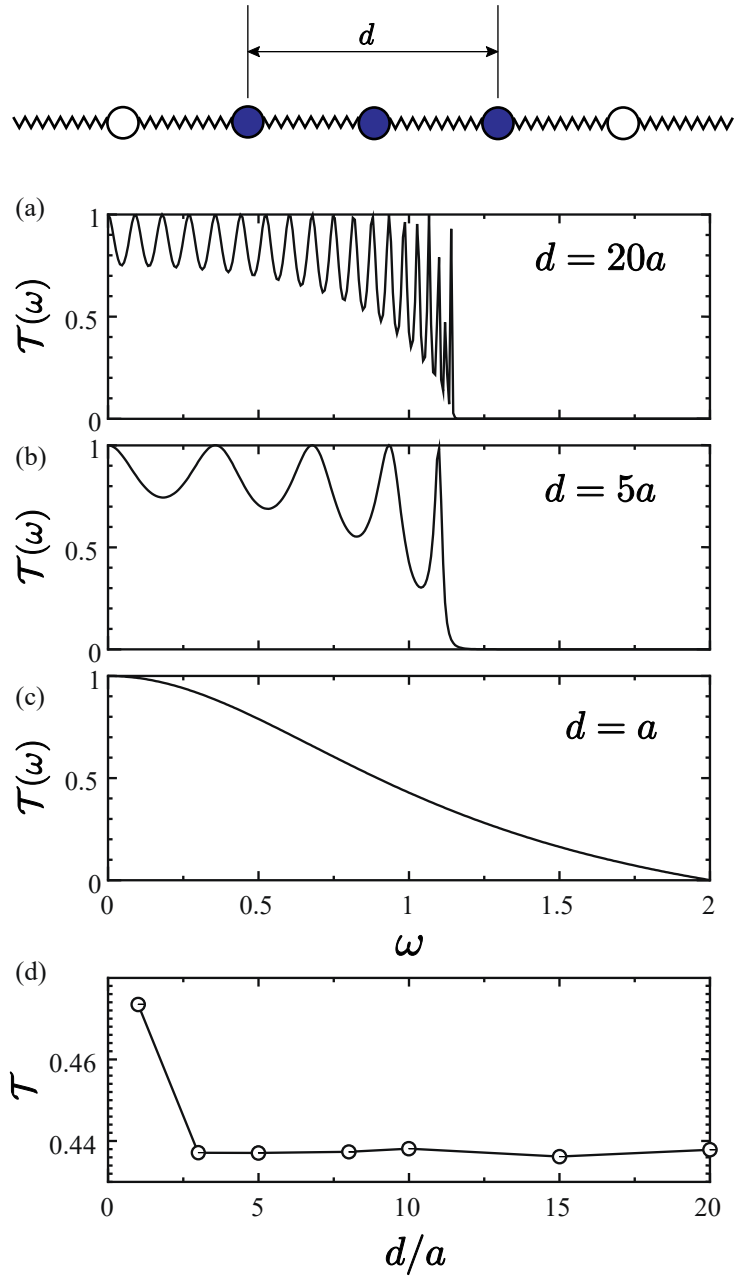

**Fig. S1** Transmission function for the scattering region with different thickness, where *d* is the thickness of scattering region, *a* is the lattice constant. (a) *d*=20*a*, (b) *d*=5*a*, (c) *d*=*a*, (d) normalized intigration of phonon transmission.

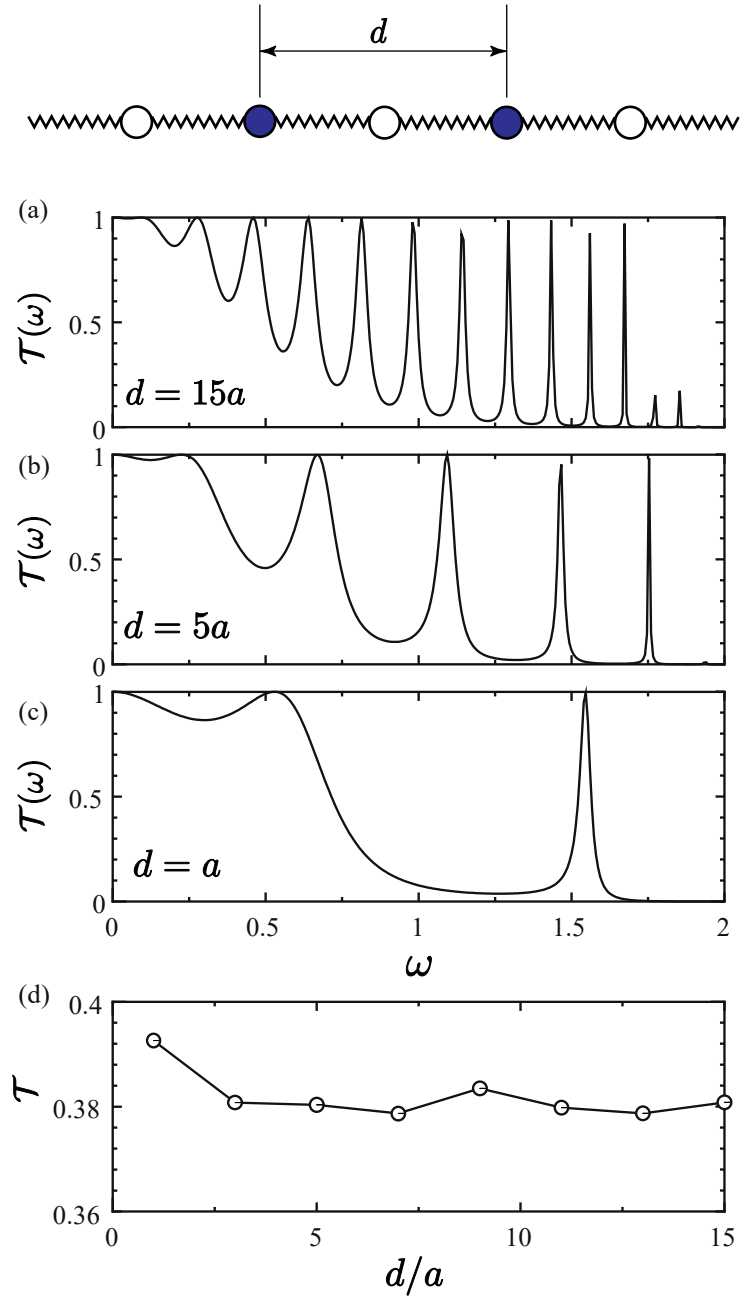

**Fig. S2** Transmission function for two interfaces with different distance, where $d$ is the distance between two interfaces, $a$ is the lattice constant. (a) $d=15a$, (b) $d=5a$, (c) $d=a$, (d) Normalized intigration of phonon transmission.